\documentstyle[aps,prl,multicol]{revtex}

\begin{document}
\input{epsf}
\draft
\tighten

\preprint{}

\title{
Individual Scatterers as Microscopic Origin of Equilibration \\
between Spin-polarized Edge Channels in the Quantum Hall regime
}
\author{Y. Acremann, T. Heinzel, and K. Ensslin
}
\address{
Solid State Physics Laboratory, ETH Z\"{u}rich, 8093 Z\"{u}rich, 
Switzerland
}
\author{E. Gini and H. Melchior
}
\address{
Institute of Quantum Electronics, ETH Z\"{u}rich, 8093 Z\"{u}rich, 
Switzerland
}
\author{M. Holland
}
\address{
Department of Electronics, University of Glasgow, Glasgow G12 8QQ, 
United Kingdom
}

\date{\today}
\maketitle
\begin{abstract}
The  equilibration length between spin-polarized edge states in the Quantum Hall regime is 
measured as a function of a gate voltage applied to an electrode on top of the edge channels.  
Reproducible fluctuations in the coupling are observed and interpreted as a 
mesoscopic fingerprint of single spin-flip scatterers which are turned on and 
off. A model to analyze macroscopic edge state coupling in terms of individual scatterers is developed, 
and characteristic values for these scatterers in our samples are extracted. For all samples investigated, 
the distance between spin-flip scatterers lies between the Drude and the quantum scattering length.
\end{abstract}
\begin{multicols} {2}
\narrowtext
 	The unique transport properties of two-dimensional electron gases 
in the integer Quantum Hall regime\cite{Klitzing80} can, to a large extent, be 
explained in a single particle picture by electronic transport via 
edge channels\cite{Halperin82}, which carry the current without dissipation over 
macroscopic distances. The Quantum Hall effect is not influenced by 
scattering between edge channels running at the same side of the 
sample. Such scattering, however, does occur and can be detected in 
various ways, for example by measuring the equilibration between edge 
channels\cite{Haug93}. In a macroscopic picture, the equilibration between two 
edge channels can be described by the coupling $P$, a macroscopic 
quantity defined by $P = (\Delta\mu-\Delta\mu') / \Delta\mu$, where 
$\Delta\mu$ and $\Delta\mu'$ are the 
differences in the electrochemical potential between the edge 
channels before and after the equilibration along a length 
$L$. The 
equilibration length $\ell_{eq}$ is then defined by the length over which $\Delta\mu$ 
is reduced to $1/e$ of its initial value, i.e. $\ell_{eq}=-L / 
\ln(1-P)$\onlinecite{Haug93}. 

This picture does not contain information on the microscopic origin 
of edge state equilibration. However, it is generally accepted that 
equilibration between spin-polarized edge channels, separated in 
energy by $g\mu_{B}B$ ($g$ is the effective electronic g-factor, $\mu_{B}$ the Bohr 
magneton, and $B$ the magnetic field) takes place via spin-orbit 
coupling\onlinecite{Haug93}, in contrast to the equilibration between edge channels 
separated in energy by the Landau gap\cite{Alphenaar90}. Impurities provide  magnetic 
field gradients in the reference frame of the moving electrons, which 
can induce scattering between edge channels of opposite spin. 
Measurements of the equilibration length over macroscopic 
distances\cite{Haren93,Herfort97} are in agreement with the values obtained from the theory 
of this spin-orbit coupling 
mechanism\cite{Komiyama92,Hirai95,Khaetskii92,Muller92,Martin91}. Other possible mechanisms 
for inter-edge state coupling, for example magnetic impurities, are 
thought to play only a secondary role in clean Ga[Al]As 
heterostructures. In a recent work, 
Polyakov\cite{Polyakov96} argues that since these spin-flip 
transitions involve a momentum transfer to the impurity potential, 
their rate must be suppressed when the disorder is smooth. He 
concludes that the spin-flip scattering is not only very sensitive to 
the local potential, but also that transitions are induced by rather 
rare fluctuations which provide a strong scattering potential.
 It is precisely this picture that we can support by the measurements 
presented here. Basically, we study the edge channel equilibration on 
a length scale short enough that fluctuations due to individual scatterers 
do not average out. By comparing the 
coupling induced by individual scatterers with the macroscopic 
coupling, we determine a spin-flip scattering length $\ell_{sf}$, i.e. the distance between scatterers 
leading to a coupling between the spin-resolved edge channels. We 
find that in all our samples, this length is smaller than 
the Drude scattering length, but significantly larger than the 
quantum scattering length. 

The samples under study are standard Ga[Al]As heterostructures (see 
Table 1, upper part, for their properties). A Hall bar geometry is 
defined by wet chemical etching, and the electron gas is accessed via 
Ni-AuGe Ohmic contacts (inset in Fig.1). Each sample contains three 
gates, two of which cross the Hall bar ( referred to as gate 1 and 
gate 2 ), and a third one (referred to as edge gate - eg ) is located in 
between gates 1 and 2. The edge gate crosses one of the mesa edges 
and extends 2 $\mu m$ onto the mesa. In order to observe a possible effect 
of individual scatterers on the edge state coupling, the length $L$ of 
the edge gate was chosen to be of the order of the equilibration length 
$l_{eq}$ between the spin-polarized edge states 1 and 
2, which was determined in previous studies to be of the order of 
$24 \mu m$. The samples were cooled down to the base temperature $T<100 mK$ of a dilution 
refrigerator. A magnetic field was applied 
perpendicular to the electron gas to establish filling factor (i.e. 
the number of occupied Landau levels) 2 in the ungated regions, and 
the longitudinal resistance $R_{xx}$ as a function of gate voltages is 
measured using a standard lock-in technique (a current of 10 nA at 
a frequency of 13 Hz). The electron densities and mobilities can be 
found in Table 1. Fig. 1 shows a measurement of $\ell_{eq}$ for sample A1163. 
First, $V_{1}$ is swept while $V_{2} = V_{eg} = +55 mV$. For $+100mV > V_{1}> 
-100mV$, $R_{xx}$  is essentially zero, indicating perfect transmission of 
both edge channels. Between $V_{1} = -100 mV$ and $V_{1}= -160 
mV$, $R_{xx}$ increases to $0.5 h/e^{2}$. The spin-down channel is now completely 
reflected. At even lower voltages ($V_{1}< -300 mV$, the spin-up channel 
gets also reflected and $R_{xx}$ approaches infinity\cite{Haug88}. We then 
sweep $V_{2}$ while keeping $V_{1}$ fixed at $V_{1}= -215 mV$, and $V_{eg}=+55 mV$. 
Again, the spin-down channel is reflected around $V_{2}=-160 mV$. However, $R_{xx}$ 
is only $0.78 h/e^{2}$, which is a measure for the edge state equilibration. This setup for 
measuring $\ell_{eq}$ has been used before by M\"{u}ller et 
al\onlinecite{Muller92}.  Using the expressions derived in Ref.\onlinecite{Muller92}, 
we find $\ell_{eq} = 18.7 \mu m$, and a coupling of $P=0.72$ for 
sample A1163. 

\begin{figure}
\centerline{\epsfxsize=3.2in\epsfbox{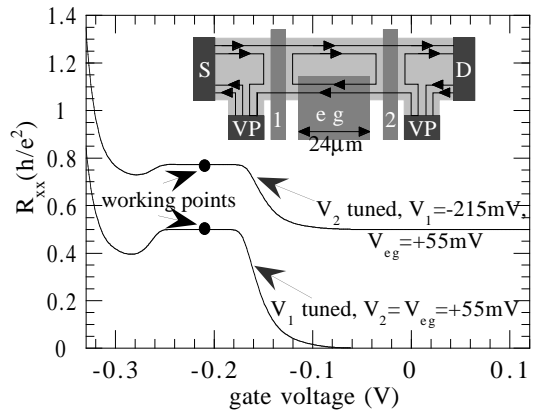}}
\caption{$R_{xx} $of sample A1163 as a function of the voltages applied to 
gates 1 and 2, respectively, in a magnetic field of $B=9.3T$ , applied 
perpendicular to the electron gas. The inset shows a scheme of the Hall bar geometry used. 
Dark areas denote Ohmic contacts used as source (S), drain (D), and 
voltage probes (VP) to the electron gas (light grey). In addition, 3 
gates  are used, gate 1, gate 2 and the edge gate (eg). Two edge channels 
are present, of which the spin-down channel gets completely reflected at gates 
1 and 2, when they are biased to the working point of -215 mV.}
\end{figure}

In the following, we focus on a voltage applied to the edge gate 
that influences the coupling between the edge channels. For these 
experiments, both gate 1 and 2 are biased at a constant voltage of 
$V_{1,2}= -215 mV$, i.e. the spin-down channel is reflected at both 
gates. If we assume that the edge channel coupling below the edge 
gate, $P_{eg}$, can be different than the edge channel coupling $P_{u}$ along 
the ungated edge,  $R_{xx}$  is determined by 
$P_{eg}$ and $P_{u}$. Using the Landauer-B\"{u}ttiker formalism\cite{Buttiker88}, one finds:

\begin{equation}
R_{xx}=\frac{h}{2e^{2}}\frac{P_{eg}+P_{u}}{P_{eg}+P_{u}-P_{eg}\cdot P_{u}}
\label{1}
\end{equation} \\
For $P_{eg}= P_{u} = P$, eq. 1 reduces to the standard expression, i.e. $R_{xx}
= (h/e^{2}) / (2-P)$\onlinecite{Muller92}.
	Fig. 2a shows $R_{xx}$ as a function of the edge gate voltage $V_{eg}$ with 
$V_{1,2} = -215 mV$, clearly indicating that the edge state coupling 
depends upon $V_{eg}$  (eq. 1). For $V_{eg}>-100mV$, the 
coupling is reduced when the gate voltage is lowered. Here, lowering $V_{eg}$ reduces the electron density 
underneath the edge gate, resulting in an increasing separation between the edge channels 
and their continuous shift away from the mesa edge. Pronounced, reproducible  fluctuations 
are visible in this regime (Fig. 2b), which we will discuss below in 
detail. At $V_{eg} = -100 mV$, a sharp drop in the coupling occurs. Using 
$P_{u} = 0.72$ from above, $P_{eg}$ drops to $0.56$ in the minimum, which means 
that $\ell _{eq}$ increases to $30\mu m$ for the edge channels below the gate. 
At $V_{eg}=-100 mV$, the spin-down channel is excluded from the gated area, as it happens 
below gates 1 and 2 (see Fig. 1) at a similar voltage, and the edge 
channel separation reaches its maximum. A further reduction of $V_{eg}$
shifts only the spin-up channel towards the edge of the gated region 
while leaving the spin-down channel basically unaffected. As below 
gates 1 and 2, the electron gas is completely depleted underneath the edge 
gate  around $V_{eg}=-300 mV$, and both edge channels run again in close proximity outside 
the region covered by the edge gate. Consequently, the coupling jumps 
back up at $V_{eg}=-300mV$.

The resistance fluctuations occur in the regime between $-100 mV \leq V_{eg} 
\leq +150 mV$ and are perfectly reproducible in each sample (Fig. 2b) 
for different sweeps of $V_{eg}$, while their 
amplitude and characteristic period depend upon the heterostructure 
used. No indication for similar fluctuations could be found as a 
function of $V_{1}$ or $V_{2}$. Typically, the fluctuation amplitude is 
of the order of $50\mit\Omega$ in our samples, and their period is of 
the order of  10 mV in $V_{eg}$. They vanish at a temperature of roughly 1K. 
We interpret these fluctuations as the effect of single impurities, each of which 
contributes an average inter-edge state coupling $q$. Since the edge gate 
voltage varies the energy and the position of the edge channels, single 
scatterers can be switched on and off. In this interpretation, the 
fluctuations can be viewed as a mesoscopic fingerprint of the 
spin-flip scatterers under the edge gate. Within our picture, we can 
obtain the average coupling induced by a single scatterer by 
analyzing the amplitude of the fluctuations, and their period gives 
information about the typical energy range over which one scatterer 
effects the edge state coupling.

\begin{figure}
\centerline{\epsfxsize=3.2in\epsfbox{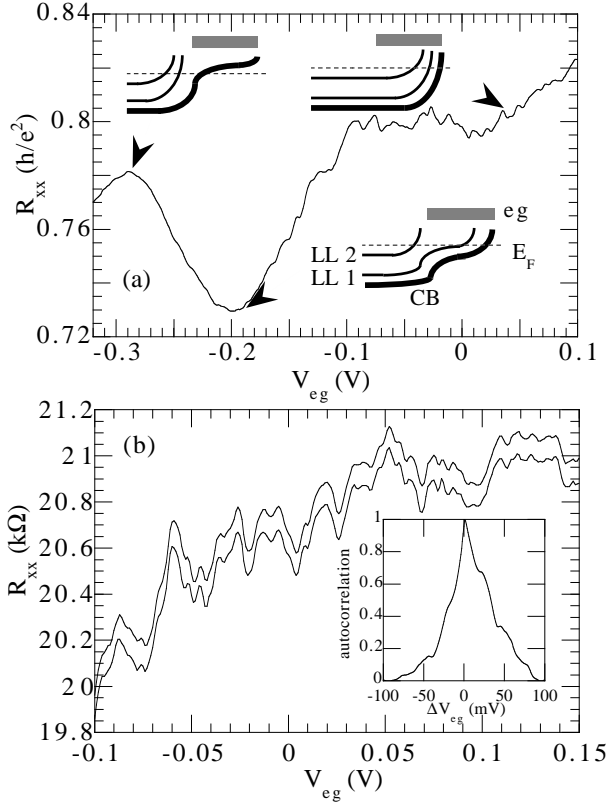}}
\caption{a) $R_{xx}$ as a function of the edge gate voltage, as measured on sample 
A1163, reflecting the variation of the inter-edge state coupling. The insets sketch 
the energy of Landau levels 1 and 2  in the corresponding regimes, indicating 
their gradual exclusion from the gated region as the edge gate voltage 
is reduced. $E_{F}$ denotes the Fermi energy, and CB the bottom of the 
conduction band. b) Reproducible fluctuations in the regime between  $-100mV \leq  
V_{eg}\leq  + 150mV$. 
The traces are offset by $100 \Omega$ for clarity. The inset shows the autocorrelation 
function as a function of the voltage shift of these fluctuations.}
\end{figure}

The average fluctuation in edge state coupling is calculated from the average fluctuation 
in $R_{xx}$ using a simple model described below. A relation between the resistance 
fluctuations and $q$ is derived (eq. 6). Once $q$ is known, the number of 
scatterers $N$ under the edge gate leading to  
equilibration can be calculated (eq. 4), and hence $\ell_{sf}$, given by $\ell_{sf} = 
L / N$, is obtained. Here, $L$ denotes the length of the edge gate.
Tuning $V_{eg}$ changes $P_{eg} $
while leaving $P_{u}$ unchanged. Assuming that $P_{eg} $ is produced 
by $N$ individual inter-edge state scatterers, each with an average coupling $q $, 
we can define $q$ as $q=1-(\Delta\mu_{i}/\Delta\mu_{i-1})$ , in analogy to the definition of 
$P$ in Ref.\onlinecite{Haug93}. Here, $\Delta\mu_{i}$ denotes the difference in the 
electrochemical potential between the two edge channels after they 
have felt i active scatterers. We will denote the coupling provided 
by N scatterers as $P_{eg,N}$. As a consequence of edge state coupling, 
the difference in the electrochemical potentials of the edge channels 
at the entrance of the edge gate $\Delta\mu_{0}$ is reduced along the gated edge, 
finally reaching $\Delta\mu_{N}$. Hence, we can rewrite the definition of the 
edge state coupling as 
\begin{equation}
P_{eg,N}=1-(\Delta\mu_{N}/\Delta\mu_{0})
\label{2}
\end{equation}
				
Using the above equation for defining $q$ iteratively, $\Delta\mu_{N}$ can be written as 
 \begin{equation}
\Delta\mu_{N}=\Delta\mu_{0}\cdot (1-q)^{N}
\label{3}
\end{equation}					
We find $N$ by inserting (3) in (2):
 \begin{equation}
N=\frac{\ln(1-P_{eg,N})}{\ln(1-q)}
\label{4}
\end{equation}						
Furthermore, we can calculate how one additional scatterer changes 
$P_{eg,N}$. Inserting (3) in (2) for $N$ and for $N+1$ scatterers, we get
 \begin{equation}
P_{eg,N+1}=P_{eg,N}+q-q\cdot P_{eg,N}
\label{5}
\end{equation}
Using (5), we can calculate how $R_{xx}$ is changed by adding one 
scatterer. For simplicity, we set $P_{eg,N} = P_{u} = P$, a reasonable 
estimate in the regime $-100 mV < V_{eg} < 100 mV$ (see Fig. 2a).
In order to obtain $q$  from the fluctuations in 
$R_{xx}$, we calculate  $\mit\Delta R_{xx}=\mit\Delta R_{xx,N+1}-\mit\Delta 
R_{xx,N}$  by inserting $P_{eg,N+1}$ and $P_{eg,N}$ from eqns. (5) and 
(2) in (1) and find  $q=\mit\Delta R_{xx}\cdot(P-2)^{2}P/
((P-1)\mit\Delta R_{xx}\cdot(2-3P+P^{2})-P\cdot h/2e^{2})$. Since the first term in the denominator is 
always small compared to the second one for our measurements, we can 
approximate
\begin{equation}
q\approx \frac{2e^{2}}{h}\cdot \mit\Delta R_{xx}\cdot(2-P)^{2}
\label{6}
\end{equation}

We have investigated 3 Ga[Al]As heterostructures 
named A1163, A577 and R2065. Samples A1163 and A577 have been grown 
by molecular beam epitaxy, MBE, with different layer sequences. In 
sample A1163, a rather high electron density is realized, while in 
sample A577, the sheet density is much lower (see Table 1). Both 
samples show Drude scattering lengths (as extracted from the 
electron density and the mobility at B=0) of the order of $3\mu m$ and 
quantum scattering lengths (obtained by fitting the Shubnikov-de Haas 
oscillations to the formula given by Ando et al.\cite{Ando82}) of the order of 
100 nm. The sample with the higher mobility (A577) also shows a 
higher equilibration length. In addition, sample R2065 has been 
grown by metal organic chemical vapor deposition (MOCVD). Both Drude 
and quantum scattering length are one order of magnitude samller in sample R2065, and $\ell_{eq}$ 
is reduced as well.
The two MBE grown samples show large $q$ values of the order of 0.02, 
while the low-mobility sample R2065, has only $q=0.0064$, but there are 
much more spin-flip scatterers. Both $N$ and $q$ scale monotonically with 
the Drude as well as on the quantum scattering length. The results 
for $q$ and the number of scatterers $N$ in each sample are summarized in 
Table 1. We find that $\ell_{sf}$ lies significantly below the Drude scattering length, but 
also significantly above the quantum scattering length in all our 
samples. We conclude that the electric field of a spin-flip scatterer 
is well below that one of a large-angle scatterer, but also well 
above that one of a small-angle scatterer. 
Our data also indicate that by increasing the number of 
spin-flip scatterers, the effective coupling per scatterer decreases 
and $\ell_{sf}$ approaches the quantum scattering length when $N$ is increased. 
When the sample gets dirtier, more scatterers contribute to spin-flip 
scattering, but each one with a reduced strength. This could mean 
that the potential gradients of different scatterers tend to cancel 
each other when the scatterer density is increased. Hence, we see 
that in order to observe the fluctuations in the edge state 
coupling,  it is essential to choose the gate length not too large 
compared to $\ell_{sf}$, in order to avoid a complete canceling of the 
resistance fluctuations.
	Furthermore, we have studied the typical energy range over 
which spin-flip scatterers are active. In the inset in Fig. 2b, the 
autocorrelation function of the fluctuations vs. $V_{eg}$ is 
shown. To translate the measured autocorrelation voltages in 
energies, we determine the lever arm a of the edge gate as $\alpha  = 
\mit\Delta E/\mit\Delta V_{eg}$ from the pinch off voltage needed for complete depletion of 
the electron gas, i.e. $\alpha  = E_{F} / (V_{0}-V_{pinch-off})$. In other words, the 
energy is changed by the Fermi energy if the voltage is varied from 
$V_{ 0}= +55mV$ down to the pinch-off voltage of $-320mV$.
	We find correlation energies $E_{C}$ of the order of $100 \mu eV$ for our 
samples A577 and R2065, and a somewhat higher value of $700 \mu eV$ for 
sample A1163.  We can estimate  $E_{C}/\ell_{sf}$ as a lower 
limit for the typical electric field of an impurity leading to edge 
state equilibration, and find values of the order of $10^{3}V/m$, 
which is roughly one order of magnitude below the characteristic 
electric field of a Drude scatterer, screened by the electron 
gas\cite{Gold88}. This number for the electric field is, however, no more than a 
crude estimate, since the effect of $V_{eg}$ on the spatial position of the 
edge sates is neglected.

	In conclusion, we have investigated the microscopic origin of edge 
channel equilibration in two-dimensional electron gases. By tuning 
the edge channels in space and energy via a gate voltage, we observe 
reproducible fluctuations of the edge state equilibration, which we 
interpret as the turning on and off of individual spin-flip 
scatterers. We have provided the framework to analyze these 
fluctuations in order to extract characteristic numbers, such as the 
edge state coupling a single scatterer induces and the average 
distance between such scatterers. Our analysis supports recent 
theoretical work, i.e. that spin-flip scattering is caused by only 
a few, but rather strong, scattering potentials. 

The authors would like to thank T. Ihn for valuable discussions. 
Financial support from the Schweizerischer Nationalfonds is gratefully acknowledged.

\begin{table*}
\begin{caption} \\
Characteristics of the samples under study. Upper part: data 
obtained from magnetotransport experiments. Lower part: numbers 
related to the inter-edge channel scattering.
\end{caption}
\begin{tabular}{cccc} 
sample& A1163&A577&R2065\\ \hline \hline
electron denstity ($10^{15}m^{-2}$)&4.5&2.7&2.0\\ \hline
mobility $(m^{2}/Vs)$&60&82&6.1\\ \hline
Drude scattering length ($\mu m$)&3.0&3.2&0.2\\ \hline
quantum scattering length (nm)&82&92&25\\ \hline \hline
equilibration length $\ell_{eq}$ ($\mu m$)&18.7&29.3&13.8 \\ \hline
average fluctuation $\mit\Delta R_{xx} (h/e^{2})$&0.0061&0.0047&0.0024 \\ \hline
correlation energy (meV)&0.7&0.1&0.2 \\ \hline
coupling $q$ per impurity&0.020&0.019&0.0064\\ \hline
number $N$ of spin-flip scatterers&63&42&270\\ \hline
spin-flip scattering length $\ell_{sf }$ (nm)&381&571&89 \\ 
\end{tabular}
\end{table*}

\end{multicols}
\end{document}